\documentclass[aps,prl,twocolumn,showpacs,superscriptaddress]{revtex4}
\usepackage{graphicx} 
\usepackage{dcolumn}  
\usepackage{bm}       
\usepackage{amsmath}
\usepackage{epsfig}

\begin{document}
\title{Tunable Plasmon Molecules in Overlapping Nanovoids}
\author{I.~Romero}
\affiliation{Donostia International Physics Center, Aptdo. 1072,
20080 San Sebastian, Spain}
\author{T.~V.~Teperik}
\affiliation{Donostia International Physics Center, Aptdo. 1072,
20080 San Sebastian, Spain} \affiliation{Institute of Radio
Engineering and Electronics (Saratov Division), \\ Russian Academy
of Sciences, Zelyonaya 38, 410019 Saratov, Russia}
\author{F.~J.~Garc\'{\i}a~de~Abajo}
\email{jga@cfmac.csic.es}
\affiliation{Instituto de \'Optica - CSIC,
Serrano 121, 20006 Madrid, Spain}
\date{\today}

\begin{abstract}
Coupled and shape-tailored metallic nanoparticles are known to
exhibit hybridized plasmon resonances. This Letter discuss the
optical properties of a complementary system formed by overlapped
nanovoid dimers buried in gold and filled with silica. This is an
alternative route for plasmon engineering that benefits from
vanishing radiation losses. Our analysis demonstrates the
possibility of designing artificial plasmon molecules on the basis
of void plasmon hybridization, which allows fine mode tuning by
varying the overlap between voids. The proposed structures could
find application to both signal processing through buried optical
elements and tunable-plasmon biosensing.
\end{abstract}
\pacs{73.20.Mf,78.67.-n,84.40.Az}
\maketitle


Plasmons in metallic particles have recently attracted considerable
attention as promising candidates to realize photonic devices,
enabling manipulation of light over sub-micron distances. The
frequency of plasmon resonances can be tuned at will by
nanoengineering the particle environment and playing with different
geometries
\cite{PRH03,paper083,paper072,paper097,NOP04,BCN05,paper114,paper115}.
The suite of available nontrivial structures includes nanoshells
\cite{PRH03,paper083}, nanorings \cite{paper072}, nanovoids
\cite{paper097}, nanorods \cite{BCN05}, and exotic polyhedral
crystallites \cite{paper115}. Plasmon hybridization originating in
intra- \cite{PRH03} and inter-particle \cite{NOP04} interactions
adds up an extra handle to fine tuning resonant mode energies,
particularly in dimers near the touching limit \cite{paper114,DN07}.

In this context, the guiding properties of nanoparticle linear
arrays have been thoroughly investigated with a view to designing
optical interconnects based upon plasmon hopping between particles
\cite{WKD01,MKA03}. However, plasmon propagation in these systems is
found to be severely damaged by both dissipation in the metal and
radiative losses. The latter are minimized by relying on small
particles ($\lesssim50$ nm), but there is a tradeoff between both
types of losses. Ultimately, absorption limits propagation to
sub-micron distances in small particles \cite{MKA03}. This problem
can be solved by using instead larger voids or dielectric inclusions
buried in metal. Actually, the radiative damping of hybridized
void-like plasmons in nanoshells decreases extremely rapidly with
increasing shell thickness \cite{paper083}. The basic unit of such
void arrays is the void dimer, to which we devote this work.

In the electrostatic limit, the mode frequencies of the particle
array $\omega_j^p$ and its complementary void array $\omega_j^v$ are
interconnected through the general, exact relation $(\epsilon_{\rm
i}+\epsilon_{\rm o})/(\epsilon_{\rm i}-\epsilon_{\rm o})=\Lambda_j$
\cite{paper025}, where $\epsilon_{\rm i}$ and $\epsilon_{\rm o}$ are
the permittivities of the material inside and outside the spherical
boundaries, respectively, and $\Lambda_j$ are eigenvalues that
depend exclusively on geometry and not on the actual dielectric
properties of the materials involved. When the media under
consideration are air and a Drude metal described by
\begin{eqnarray}
\epsilon(\omega)=1-\frac{\omega_p^2}{\omega(\omega+i\eta)},
\label{Drude}
\end{eqnarray}
one obtains the sum rule $(\omega_j^p)^2+(\omega_j^v)^2=\omega_p^2$
\cite{AER96}. Therefore, the analysis of electrostatic void arrays
is fully contained in the electrostatic particle array. We will
however consider relatively large voids yielding longer propagation
distances. Then, retardation effects become dominant for sizes above
the skin depth ($\sim 25$ nm at visible and NIR frequencies), and
therefore a separate analysis becomes necessary.

In this work, we describe the optical properties of plasmon
molecules formed in overlapping nanovoids buried in metal. Similar
to particle dimers \cite{NOP04}, large shifts of hybridized modes
are observed, controlled by varying the degree of overlap, while
significant field enhancement takes place near the junction region.
But unlike particle dimers, void dimers do not present a singular
transition near touching conditions. More precisely, the unphysical
mode observed in the separate particle dimer \cite{paper114},
characterized by a net induced charge sitting in each particle,
becomes now physical for the voids, although the inter-void
interaction is drastically reduced in non-overlapping systems due to
the skin-depth effect.

We rigorously solve Maxwell's equations in the frequency domain
$\omega$ for overlapping void dimers filled with silica using the
boundary element method (BEM) \cite{paper030paper070}, in which
surface charges and currents are introduced to match the fields
across dielectric boundaries, thus resulting in a set of
surface-integral equations that is transformed in turn into a
linear-algebra problem upon discretization of the integrals. The NIR
permittivity of gold is modeled for simplicity through Eq.\
(\ref{Drude}) with parameters $\omega_p=7.9$~eV and $\eta=0.09$~eV
\cite{KMK02}, while we take $\epsilon_{\rm i}=2$ for silica.

Fully buried voids are unaccessible to external radiation, so light
scattering cannot be employed to characterize them. Instead, we
investigate the photonic local density of states (LDOS) in the
silica region, defined as the combined local intensity of all
eigenmodes of the system under consideration \cite{FMM04}. With this
definition, the LDOS $\rho$ projected along the direction of an
atomic-decay transition-dipole ${\bf d}$ is related to the decay
rate as $\Gamma=(4\pi^2\omega d^2/\hbar)\rho$. In practical terms,
the LDOS can be calculated from the field scattered by the void
boundaries, projected along the dipole, according to
\begin{eqnarray}
\rho=\frac{\omega^2\sqrt{\epsilon_{\rm
i}}}{3\pi^2c^3}+\frac{1}{2\pi^2\omega d^2}{\rm Im}\left\{{\bf
d}^*\cdot{\bf E}^{\rm scat}\right\}, \nonumber
\end{eqnarray}
where the first term on the right hand side is the silica LDOS.
Here, we consider projection directions parallel and perpendicular
to the dimer axis and give the LDOS normalized to its value in
silica.

\begin{figure}
\includegraphics{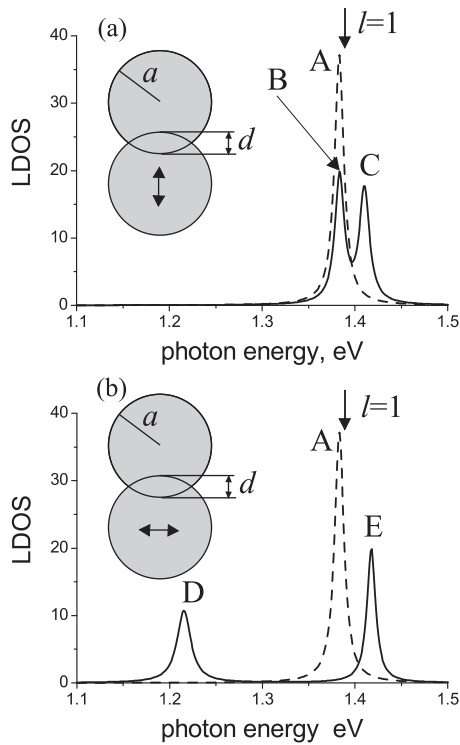}
\caption{\label{first} Plasmon modes in overlapping nanovoids. The
photon local density of states (LDOS) is calculated inside two
overlapping silica inclusions in gold (solid curves) and projected
along directions (a) parallel and (b) perpendicular to the dimer
axis (see insets). The inclusions are spheres of radius $a=240$~nm.
The overlapping distance is $d=60$~nm. The LDOS of a single
inclusion is shown as dashed curves. The vertical arrows mark the
dipole plasmon energy of a single inclusion obtained from
Eq.~(\ref{vwwoidmodes}) with $l=1$.}
\end{figure}

\begin{figure*}
\includegraphics{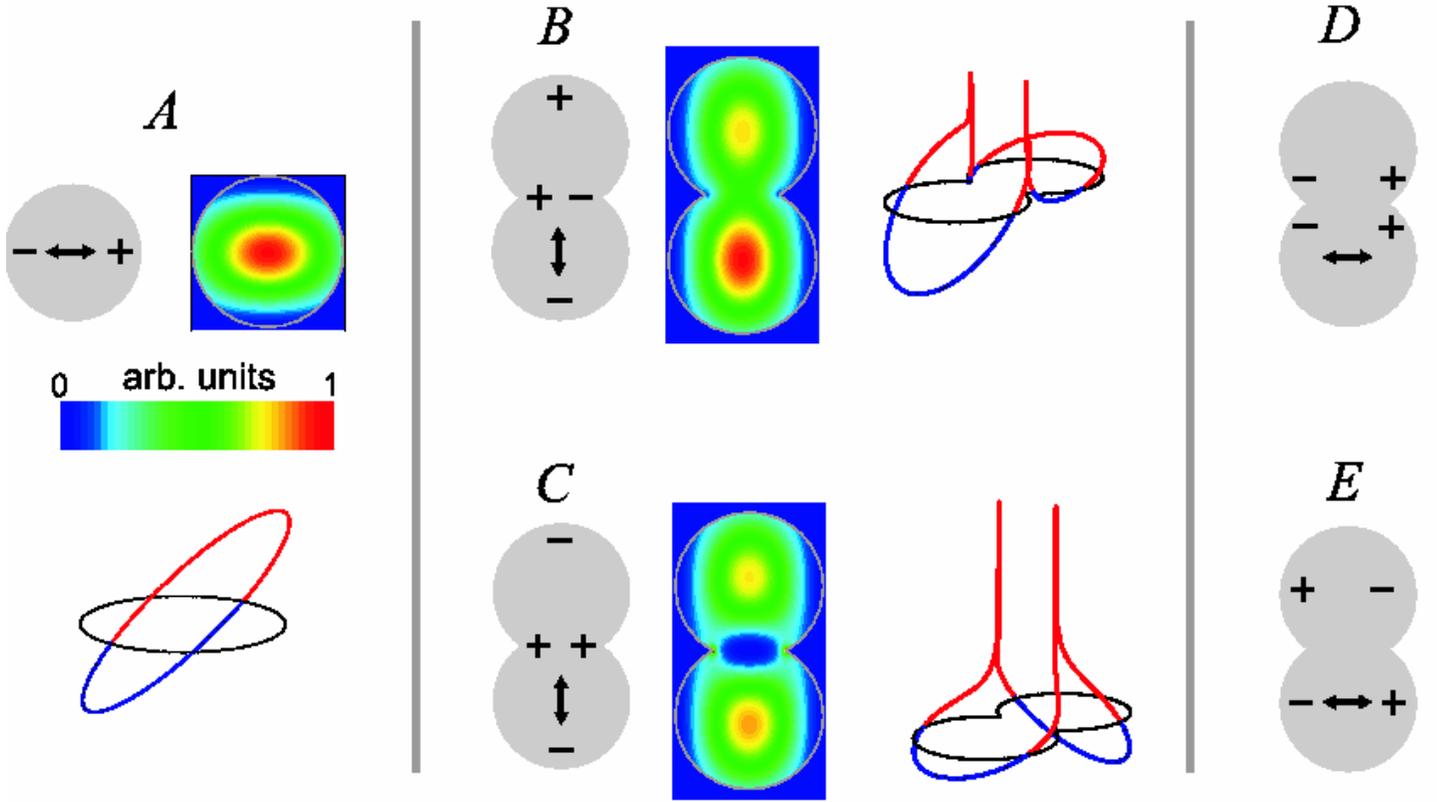}
\caption{\label{second} Plasmon mode maps corresponding to peaks A-E
in Fig.\ \ref{first}. The contour plots show the square of the
induced electric field produced by dipoles as indicated by the
double arrows. Schemes of the charge distribution are shown to the
left of each contour plot. The induced surface charge is shown as
well, with negative and positive values represented in blue and red,
respectively.}
\end{figure*}

\begin{figure*}
\includegraphics{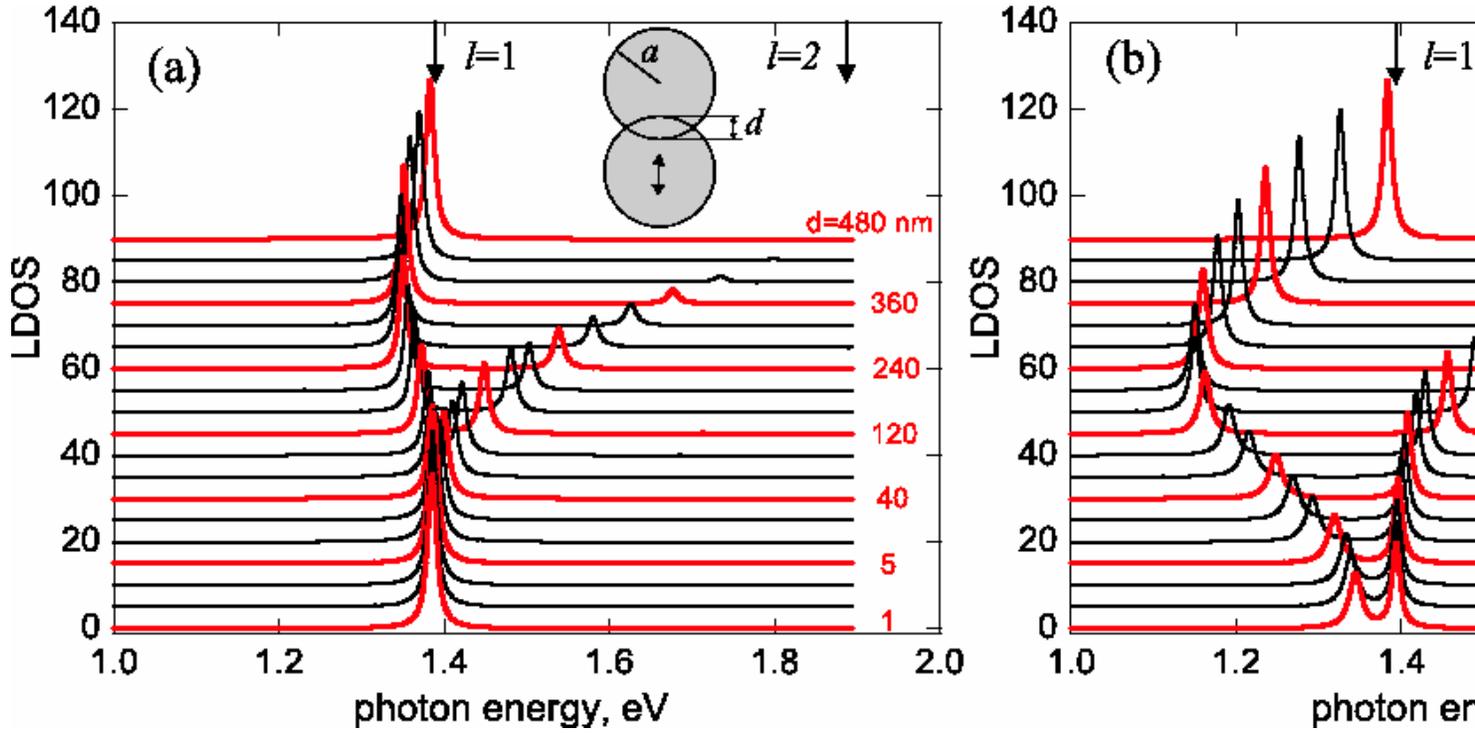}
\caption{\label{third} Evolution of void-dimer plasmon energies with
overlapping distance $d$. The modes are probed through the LDOS at
the center of one of the voids, projected along the dimer axis in
(a) and perpendicular to that axis in (b). The void radius is
$a=240$~nm. The vertical arrows mark the energy of the dipole
($l=1$) and quadrupole ($l=2$) plasmon modes of a single inclusion
obtained from Eq.~(\ref{vwwoidmodes}).}
\end{figure*}

\begin{figure}
\includegraphics{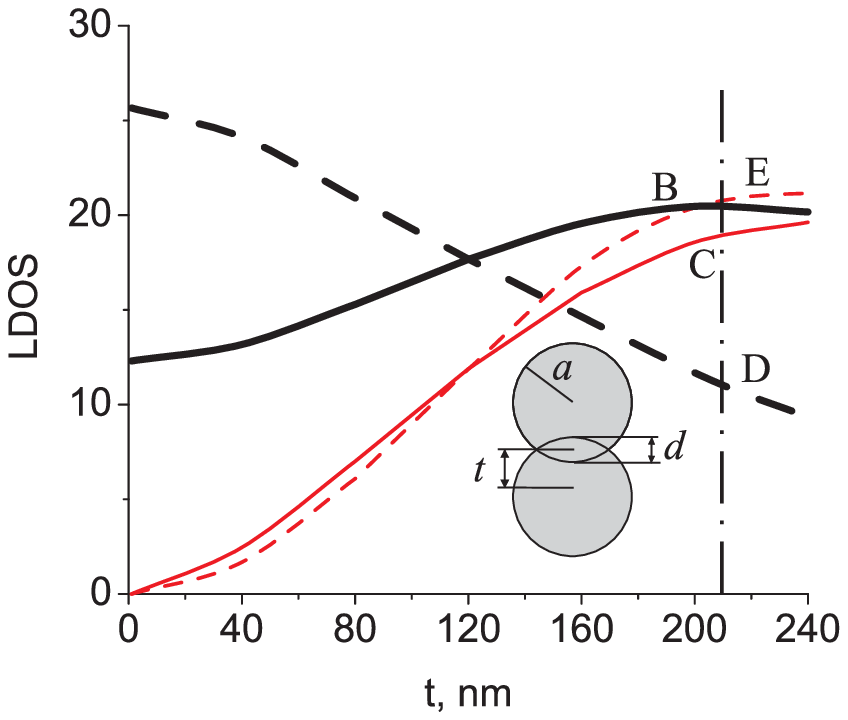}
\caption{\label{fourth} LDOS calculated for two overlapping silica
inclusions in gold as a function of position $t$ along the dimer
axis. The LDOS is projected along parallel (solid curves) and
perpendicular (dashed curves) directions relative to that axis.
Thick and thin curves correspond to low- and high-energy resonances,
respectively. The void radius is $a=240$~nm and the overlapping
distance is $d=60$~nm.}
\end{figure}

We show in Fig.\ \ref{first} the LDOS at the center of one of the
silica-filled voids of a dimer formed by spheres of radius $a=240$
nm with d=60 nm overlap (solid curves). For comparison, the
isolated-void LDOS is also represented (dashed curves), exhibiting a
prominent feature corresponding to the $l=1$ electric Mie resonance
in agreement with the analytical expression (see arrows) \cite{W69}
\begin{equation}
\epsilon_{\rm o}h_l^{(1)}(\rho_{\rm o})\frac{d[\rho_{\rm
i}j_l(\rho_{\rm i})]}{d\rho_{\rm i}}= \epsilon_{\rm i}j_l(\rho_{\rm
i})\frac{d[\rho_{\rm o}h_l^{(1)}(\rho_{\rm o})]}{d\rho_{\rm o}},
\label{vwwoidmodes}
\end{equation}
where $\rho_j=\sqrt{\epsilon_j}\omega a/c$, $j_l$ and $h_l^{(1)}$
are spherical Bessel and Hankel functions, and $l$ is the orbital
momentum number. The small mismatch between BEM calculations and the
analytical mode frequency originates in non-resonant contributions
to the dipole-induced field of the former.

The $l$-plasmon mode of the single void is $2l+1$ degenerate, and
one can expect partial lifting of this degeneracy due to void
overlap. This is clearly observed in Fig.\ \ref{first}. The
resulting modes have $m=0$ and $m=\pm 1$ azimuthal symmetry in
Figs.\ \ref{first}(a) and (b), corresponding to transition dipoles
oriented parallel and perpendicular to the dimer axis, respectively.
The latter is doubly degenerate, so that the overall number of 6
modes is preserved after inter-void interaction is switched on. The
dimer modes are hybridized and energy split, but the splitting is
significantly larger for the transverse low-energy feature D. The
origin of this can be traced back to strong interaction within the
inter-void opening, which is made possible for co-directional
transverse orientations of the exciting dipoles and has similar
nature as gap modes in particle dimers \cite{NOP04}. In a related
context, this gives rise to rim modes localized at the opening of
voids partly buried in flat metal surfaces \cite{paper128}.

This intuition is supported by the induced-field distributions shown
in Fig.\ \ref{second} for features A-E of Fig.\ \ref{first}. See for
instance the large increase in field intensity near the void
junction region associated to peak D. This is accompanied by
significant piling up of induced charge, as shown in the graph to
the right of the contour plot. However, peak E has vanishing
intensity in the voids opening due its quadrupolar symmetry, and
therefore it suffers a rather moderate shift.

These results suggest the possibility to design plasmon molecules of
tailored frequency and symmetry. For example, a quadrupole plasmon
mode can be obtained in the dimer near the single-void dipole mode
frequency (E in Figs.\ \ref{first} and \ref{second}). More
sophisticated plasmon modes should also be possible by combining
several voids, with the dimer acting as a building block.

In contrast to metallic-particle dimers, in which plasmons interact
through a dielectric environment \cite{paper114}, our void plasmons
are coupled only through the region of overlap. Therefore, plasmon
mixing critically depends on the overlap distance $d$ and can be
significantly reduced in buried geometries, thus simplifying the
understanding of the resulting plasmon molecule. In particular,
modes B, C, and E are reliably described in terms of dipole-dipole
interaction, whereas mode D, which lies further away from the
single-void dipole resonance, is dominated by the rim-like mode,
involving a large degree of mixing of higher-order modes.

The full $d$ dependence of the void dimers is presented in Fig.\
\ref{third}. Several characteristic features are clearly resolved in
the evolution of the plasmon modes:

(i) Mode splitting is consistently larger for the transverse
orientation of the dipole, in good agreement with the conclusions
extracted from Figs.\ \ref{first} and \ref{second}.

(ii) As anticipated above, the interaction between voids switches on
smoothly between non-touching and overlapping configurations, in
contrast to the singular transition observed for particle dimers
\cite{paper114}.

(iii) The low-energy dipole modes evolve smoothly towards the
single-void dipole when the overlap is complete ($d=2a$). An initial
red shift after touching is followed by a blue shift as $d$
increases. In the rim-like mode [see D in Figs.\ \ref{first} and
\ref{second}, and low-frequency features in Fig.\ \ref{third}(c)], a
maximum red shift is produced at an optimum distance resulting from
the compromise between increasing mode volume as $d$ becomes smaller
(this lowers the {\it kinetic energy} of the mode) and increasing
inter-void interaction as $d$ increases (this lowers the mode energy
due to inter-mode attraction).

(iv) The high-energy quadrupolar modes march towards the single-void
quadrupole as $d$ increases and the quadrupole-mode LDOS becomes
zero at the center of a single-void.

(v) A weak transition resonance is observed in the vicinity of 1.7
eV [Fig.\ \ref{third}(b)], originating in higher-order void
multipoles that are the counterpart of the left-most rim-like mode.

The plasmon widths determine how well the modes are defined. In this
sense, the 1.7 eV resonance just noted is a well-defined one. In
fact, the width of all our reported plasmons arise from metal
absorption, so that modes associated to fields mainly contained in
the silica and characterized by minimum overlap with the metal will
have longer lifetimes. This is clear when comparing modes D and E in
Figs.\ \ref{first}(b) and \ref{second}: D is broader because it has
larger intensity in the metal (see region near the void junction).
However, much of the field intensity lies in the silica, and that is
why the mode widths are considerably smaller than $\eta$ in Eq.\
(\ref{Drude}).

Finally, the location of the probing dipole determines the actual
degree of coupling to the plasmons. This is illustrated in Fig.\
\ref{fourth}, which shows the evolution of the plasmon LDOS along
the dimer axis for features B-E of Fig.\ \ref{first}. Interestingly,
only the low-energy hybrid plasmons (B and D), which have the
original dipole symmetry, contribute to the LDOS at the dimer center
(thick curves). The high-energy modes have an overall quadrupolar
character and can only be excited by dipoles displaced with respect
to the dimer central region.

In summary, we have investigated the optical properties of two
overlapping silica inclusions embedded in gold and found hybridized
plasmon modes, the mixing of which can be effectively controlled by
changing the degree of geometrical overlap. This opens up the
possibility to realize radiative-free plasmon molecules through
dielectric inclusions buried in metal. Large field enhancements are
observed on resonance near the overlap region, which can find
application to surface-enhanced Raman spectroscopy and other forms
of biosensing. The void dimer constitutes the basic building block
of buried optical circuits for future on-chip interconnects and
switching elements. In contrast to particle arrays, void arrays will
benefit from minimum cross-talk across the metal and vanishing
radiative losses, thus allowing a step forward in nanoscale
integration.

This work has been supported in part by the Spanish MEC
(NAN2004-08843-C05-05) and by the EU (NMP3-CT-2004-500252 and
STRP-016881-SPANS). T.V.T. acknowledges support from the Russian
Academy of Sciences, Russian Foundation for Basic Research (Grants
05-02-17513, 07-02-91011, and 06-02-81007).

\bibliographystyle{apsrev}

\end{document}